\documentstyle[seceq,preprint,epsf]{jpsj}
\title
{Theoretical Description of Resistive Behavior near a Quantum 
Vortex-Glass Transition}
\author{Hideharu Ishida and Ryusuke Ikeda}
\inst{Department of Physics, Kyoto University, Kyoto 606-8502}
\recdate{June 4, 2001}

\abst{
Resistive behaviors at nonzero temperatures ($T > 0$) reflecting a quantum 
vortex-glass (VG) transition (the so-called field-tuned 
superconductor-insulator transition at $T=0$) are studied based on a quantum 
Ginzburg-Landau (GL) action for a s-wave pairing case containing 
microscopic details. The ordinary dissipative dynamics of the pair-field is 
assumed on the basis of a consistency between the fluctuation conductance 
terms excluded from GL approach and an observed negative magnetoresistance. 
It is shown that the VG contribution, $G_{\rm vg}$, to 2D conductance becomes 
insensitive to $T$ at an {\it apparent} VG transition field $B_{\rm vg}^*$ 
defined at experimentally accessible temperatures but depends on the repulsive 
electron-electron interaction, and that, only in the dirty limit 
with no electron-repulsion, it takes a universal value at low $T$. 
Available resistivity data near $B_{\rm vg}^*$ are explained based on our 
results, and an extension of the theory to 3D case is 
briefly discussed. 
}

\kword{Superconductor-Insulator Transition, Vortex-Glass Transition, 
Vortex States}

\begin{document}

\sloppy
\maketitle

\section{Introduction}

The issue of quantum superconductor-insulator (FSI) transition induced by an applied field in disordered superconductors has become 
one of long-standing problems in 
condensed matter physics. Most of resistivity data\cite{HP,Okuma1,Gan1,Okuma2,Chicago} suggestive of an FSI 
transition have been discussed based on a scaling hypothesis postulated by 
Fisher\cite{MPAF} on resistive behaviors at nonzero temperatures ($T > 0$) accompanying a 2D vortex-glass (VG) transition at $T=0$. However, no theoretical 
calculation justifying his scenario on resistive behavior at $T>0$ has been 
reported so far. In fact, recent experimental studies\cite{Valles,Palo} 
have often led 
even to an argument {\it against} the 2D VG ordering, 
i.e., superconducting ordering in $B \neq 0$, although theoretically the 
2D quantum VG phase will exist more likely than the 3D VG phase\cite{FFH} 
does at $T > 0$ because the former is formally similar to the thermal 3D glass 
phase due to line defects\cite{NV,RI1} which certainly exists at $T > 0$. 

Fisher's scenario is based on the phase-only approximation of a GL action. 
The phase fluctuation in a phase-only model for {\it homogeneously} 
disordered (i.e., nongranular) superconductors has no dissipative dynamics, 
and hence, a {\it nondissipative} dynamics was assumed there.\cite{MPAF} 
Further, since such a phase-only action is essentially equivalent 
to the quantum boson action at 
low energy, possible vortex phases were identified with the corresponding ones 
of the boson system. However, as in the description of thermal 
vortex phase diagram at nonzero temperatures,\cite{RI2} 
a more microscopic GL approach 
in which the pair-field yields a dissipative dynamics should be able to 
explain physical properties systematically even 
at low temperatures as far as a nontrivial intermediate phase\cite{com1} 
does not 
occur as a consequence of a phase-only model valid for nongranular films. 
For instance, the so-called bose insulating state\cite{HP,Okuma1} 
will correspond to the quantum vortex liquid regime\cite{RI3} with insulating 
GL fluctuation conductivity. 

Actually, there are several reasons why the Fisher's argument on the resistive 
behavior near a FSI transition should be reconsidered theoretically. First, 
the argument\cite{MPAF} on a universal conductance value at the VG transition 
field, $B=B_{\rm vg}$, and on the scaling behavior at $T>0$ of resistance 
is based on consideration and calculations of the fluctuation 
conductance $G_s(T=0$, $\omega \to 0)$ in the disorder-tuned case 
at\cite{FGG} $B=0$, where $\omega \to 0$ means taking the dc limit. However, 
at a $T=0$ criticality, this conductance need not be equal 
to the quantity\cite{B=0} $G_s(T \to 0$, $\omega=0)$. 
It is the latter which is 
measurable in real experiments, while it is 
the former which is expected to take a universal value as a consequence of a 
quantum continuous transition.\cite{FGG} Further, 
the temperature range over which the presence of 
a $T$-independent conductance is suggested is often\cite{HP,Gan1} 
broad so that the dissipative dynamics, neglected in ref.6, of the pair-field 
does not seem to be 
negligible, and thus, it is unclear to what extent taking the $T \to 0$ limit 
is essential. In addition, the assumption\cite{MPAF} that the derivation of 
$G_s(T=0$, $\omega \to 0)$ at $B=0$ is applicable to the nonzero field case is 
not justified, because the VG critical properties of response quantities are 
quite different from those for the $B=0$ superconducting 
transition. For example, the 3D VG phase due to point-like quenched disorder 
has no static Meissner response to any disturbance of magnetic field, 
and hence 
the VG transition is not\cite{RI4} 
accompanied by a divergent diamagnetic susceptibility 
in contrast to the case of the normal-Meissner transition. 
It means that it is not\cite{RI1,RI4} justified at all to apply\cite{MPAF} 
the scaling argument on transport quantities in the normal-Meissner transition 
to the FSI transition case. 

Previously, one of the present authors has pointed out\cite{RI3} 
the possibility that the intervening metallic behavior at low $T$ may imply 
the vortex flow conductance taking a nearly universal value along the 2D 
melting line $B_m(T)$ of the disorder-free vortex solid, if $B_m(T)$ is 
insensitive to $T$ in the {\it quantum} regime. Such a quantum melting line 
insensitive to $T$ is satisfied in the dirty limit\cite{Parks} for a s-wave pairing 
and neglecting a repulsive interaction between electrons. Further, the 
$B_m(0)$ is implicitly assumed in this proposal\cite{RI3} to lie above a $T=0$ VG 
transition field $B_{\rm vg}$. As demonstrated elsewhere,\cite{RI5} however, 
$B_{\rm vg}$ calculated in a mean field approximation in terms of microscopic 
parameter in the $s$-wave dirty limit\cite{Parks} seems to, contrary to the observed trend,\cite{HP,Okuma2,Gold} increase with increasing disorder measured 
by $(E_{\rm F} \tau)^{-1}$ due primarily to the $\tau$-dependence of the 
corresponding $H_{c2}(0) \equiv H_{c2}^d(0) = 0.56 \phi_0 
T_{c0}\tau/l^2 \propto \tau^{-1}$, 
where $E_{\rm F}$ is the fermi energy, $l=v_{\rm F} \tau$ 
the mean free path, $T_{c0}$ the mean field transition temperature in zero field and in clean limit, and $\phi_0$ the flux quantum. 
This is not {\it physically} surprising, because not only 
the thermal fluctuation but also the vortex pinning are expected to be enhanced with increasing the microscopic disorder. 
Hence, an inclusion of the electronic 
interplay between the repulsive interaction and disorder seems to be 
necessary in order to get both $B_{\rm vg}$ and $H_{c2}(0)$ decreasing with 
increasing disorder or decreasing the 
film thickness.\cite{Gold} Actually, we have judged\cite{RI5} 
that a situation in which 
the 2D melting field $B_m(0)$ lies above $B_{\rm vg}$ will seldom happen 
and hence that a contribution to the conductivity arising from vortex pinning 
effects may not be negligible in explaining consistently the flat resistance 
curve (the intervening metallic behavior). In ref.11, the VG fluctuation term 
$G_{\rm vg}(T=0, \omega \to 0)$ of conductance was considered as a byproduct of transport properties near the 3D thermal glass transition due to line 
disorder and was argued to be a nonuniversal quantity in general. 
However, there, no microscopic details were taken into account, and the 
measurable quantity $G_{\rm vg}(T \to 0, \omega=0)$ was not 
examined.\cite{RI1} 

In this paper, we examine the vortex-glass contribution $G_{\rm vg}(\omega=0)$ 
to the dc conductance for a current perpendicular to an applied field at {\it 
nonzero} temperatures based on the microscopic study\cite{RI5} of quantum GL 
action for the $s$-wave pairing case. As demonstrated in ref.19, the 
coefficient of dissipative term has a remarkable $T$-dependence rather in 
the vicinity of $T=0$, implying that the premise in previous works\cite{MPAF} 
that one may start from a bosonic model at $T=0$ is not justified in the 
present issue. In $\S 2$, 
$G_{\rm vg}(T > 0, \omega=0)$ in the quantum critical region around 
and at the critical field $B_{\rm vg}$ is examined on a general ground. 
As in ref.19, the ordinary dirty limit neglecting an interplay between 
an electron-electron interaction and disorder will be called merely as the 
dirty limit, and the dynamics of the pair-field at low frequencies 
is assumed according to ref.19 to remain dissipative even 
in $T \to 0$ limit. Then, we find that 
$G_{\rm vg}(B_{\rm vg})$ becomes a nonuniversal constant depending on a 
strength of the electron-repulsion but that, in the dirty limit with no 
electron-repulsion, it unusually becomes a {\it universal} constant 
independent of material parameters. In $\S 3$, 
it is pointed out that, in $B>B_{\rm vg}$, the "bosonic" contribution 
including $G_{\rm vg}$, arising from the GL action, to the conductance 
approaches zero in $T \to 0$ limit and 
that a negative magnetoresistance of a superconducting origin\cite{Okuma2} at lower 
temperatures is provided by "fermionic" fluctuation contributions, 
such as the Maki-Thompson fluctuation term, excluded from the GL description. 
Based on these results, the resistive behavior near 
$B_{\rm vg}$ is discussed in $\S 3$ and compared with existing key resistive 
data in thin films. In $\S 4$ we comment on extensions of the present theory 
to 3D systems in low $T$ limit. 

\section{2D VG Conductance in the Quantum Critical Regime}

Since an FSI transition usually occurs far from the region near the 
zero-field superconducting transition where the vortex-pair excitations play 
essential roles, a high $B$ approximation will be invoked in which the 
pair-field in any static vortex state is described in terms of the lowest 
Landau level (LLL) modes which do not accommodate vortex-pair excitations. 
If first neglecting the random potential terms leading to the vortex pinning 
effects, the 2D quantum GL action on LLL fluctuations $\Psi$ of the pair-field 
takes the form\cite{RI3} 
$$S_{\rm unp} = \int d^2r \biggl[ \beta \sum_\omega 
( \mu(0) + \gamma |\omega| ) |\Psi_\omega({\bf r})|^2 + {{U_4} \over 2} 
\int_0^\beta du |\Psi({\bf r}, u)|^4 \biggr], \eqno (2.1)$$
where $\gamma$, $U_4 > 0$, $\beta=1/k_{\rm B} T$, $\Psi({\bf r}, u) 
= \sum_\omega \Psi_\omega({\bf r}) e^{-{\rm i} \omega u}$, $\omega$ is a 
Matsubara frequency for bosons, and the fact that the squared gauge-invariant 
gradient ${\bf Q}^2 = (-{\rm i} \nabla + {2\pi/\phi_0} {\bf A})^2$ is replaced 
by the factor $r_B^{-2} = 2 \pi B/\phi_0$ after operating any LLL 
eigenfunction was used. The mean field $H_{c2}(T)$-line is defined as 
$\mu(0)=0$. In this paper, we focus on the 
temperature range defined by $T < T_{cr}^{\rm mf}$, where\cite{RI5} 
$$T_{cr}^{\rm mf} \simeq 0.15 T_{c0} B/H_{c2}^d(0). \eqno(2.2)$$ 
As is explained later, this temperature scale arises from the denominator 
of diffusion propagators, and in $T < T_{cr}^{\rm mf}$, $\mu(0)$ and $U_4$ 
become $T$-independent on cooling, while the $T$-dependence of 
$\gamma$ depends remarkably on the presence of an electron-electron repulsive 
interaction. 

The random potential terms of GL action were studied in ref.19. At high $T$ ($< T_{c0}$) and low $B$ ($< H_{c2}(0)$), they may be represented simply in terms of a single random $T_c$ term, i.e., as a local potential form, while, in low $T$ and high $B$ case of our interest, they become spatially nonlocal reflecting 
the fact that the only microscopic scale measuring the spatial variations of $\Psi$ in high $B$ and in 2D is the averaged vortex spacing $r_B$. 
The {\it replicated} GL action within LLL arising after the random-averaging is of the form\cite{RI5} 
$$S^n_{\rm p} = \sum_\alpha \biggl[ \sum_\omega \biggl( (\mu(0) 
+ \gamma |\omega|) \sum_p |\varphi_0^{(\alpha)}(p, \omega)|^2 + {{U_4} \over 
{4 \pi r_B^2 \beta}} N_v^{-1} \sum_{\bf k} \rho^{(\alpha)}({\bf k}, \omega) 
\rho^{(\alpha)}(-{\bf k}, -\omega) \biggr)$$
$$ - \sum_{\alpha'} {{U_p} \over {4 \pi r_B^2 N_v}} \sum_{\bf k} 
f_{00}(k^2) \, \rho^{(\alpha)}({\bf k}, 0) \rho^{(\alpha')} (-{\bf k}, 0) 
\biggr], \eqno(2.3)$$ 
where $\alpha$ and $\alpha'$ are replica indices, $N_v$ is the number of 
field-induced vortices, and 
$\rho^{(\alpha)}({\bf k}, \omega)$ is the Fourier transform 
of $|\Psi^{(\alpha)}({\bf r}, \tau)|^2$ and expressed by 
$$\rho^{(\alpha)}({\bf k}, \omega) = \sum_{p, \omega_1} e^{{\rm i} p k_x - 
{\bf k}^2/4} \, \varphi_0^{(\alpha) \, *}(p-k_y/2, \omega_1) \, 
\varphi_0^{(\alpha)}(p+k_y/2, \omega_1+\omega), \eqno(2.4)$$ 
where $\Psi({\bf r}) = \sum_p \varphi_0(p) u_{0,p} ({\bf r})$ with LLL 
eigenfunction $u_{0,p}({\bf r})$ in a Landau gauge. Further, the length 
scales and $\Psi$ were rescaled, respectively, in the manners, ${\bf r}/r_B 
\to {\bf r}$ and $\beta^{1/2} \Psi \to \Psi$. The function $f_{00}(k^2)$ 
is positive and a regular function of $k^2$ (and also of ${\bf k}/|{\bf k}|$ 
when the Fermi surface is anisotropic) and, at least in the dirty limit, 
independent\cite{RI5} of material parameters such as $l$. 
Although its detailed functional form is not known even in 
the dirty limit, just the property that the wavenumber ${\bf k}$ in $f_{00}$ 
is entirely scaled by $r_B$ becomes essential in examining the critical 
conductance. 
The familiar random $T_c$ model corresponds to the specific case in which the 
$k^2$-dependences in $f_{00}(k^2)$ are neglected. 
Although this nonlocality in $f_{00}$ is 
safely negligible at high $T$ and low $B$ where the field is measured through the ratio $\xi_0^2/r_B^2$ with GL coherence length in dirty limit $\xi_0 \simeq \sqrt{v_{\rm F} l/T_{c0}}$, as already 
mentioned, it cannot be neglected in high $B$ and low $T$. 
As well as $U_4$, the bare pinning strength $U_p$ can be regarded as 
being $T$-independent\cite{RI5} 
in $T < T_{cr}^{\rm mf}$. 

Until reaching eq.(2.27), we assume in this section the GL coefficients $\gamma$, $U_4$, and $U_p$ to be $T$-independent as if the GL action is the expression 
at $T=0$. First, to illustrate properties of VG fluctuation, we invoke a systematic loop (or $1/M$) expansion and focus\cite{RI1,RI5} on its lowest order 
($M = \infty$) terms by, as a mathematical tool, 
assuming that the complex scalar pair-field $\Psi$ has $M$-flavors. Up to the 
lowest order in $U_p$, the random-averaged propagator ${\cal G}_0$ of LLL 
fluctuation $\varphi_0^{(\alpha)}$ in this case satisfies\cite{RI5} 
$$({\cal G}_0(\omega))^{-1} = \mu(0) + \gamma|\omega| + {{U_4} 
\over {4 \pi r_B^2} \beta} \sum_\omega {\cal G}_0(\omega) 
- \Delta^{(R)}_0 {\cal G}_0(\omega). \eqno(2.5)$$
Here the factor $\Delta^{(R)}_0$ is a coefficient of a 
renormalized pinning vertex off-diagonal in the 
replica indices and is given by 
$$\Delta^{(R)}_0 = {{U_p} \over {2 \pi r_B^2}} N_v^{-1} \sum_{\bf k} 
e^{-{\bf k}^2/2} f_{00}({\bf k}^2) (1 + \sigma_{\rm vg} 
e^{-{\bf k}^2/2})^{-2}, \eqno(2.6)$$
where $\sigma_{\rm vg} = (U_4/2 \pi \beta r_B^2) \sum_\omega 
{\cal G}_0^2(\omega)$. 
The solution of eq.(2.5) is easily found to have the form 
$${{{\cal G}_0(0)} \over {{\cal G}_0(\omega)}} = 1 
+ {\cal G}_0(0) {\gamma \over 2}|\omega| + {{2 t_{{\rm vg},0}^{-1} {\cal G}_0(0) \gamma |\omega| s(|\omega|)} \over {1 + \sqrt{1 + 4 t_{{\rm vg},0}^{-2} {\cal G}_0(0) \gamma |\omega| s(|\omega|)}}}, \eqno(2.7)$$
where $s(|\omega|) = 1-t_{{\rm vg},0}/2 + \gamma|\omega| {\cal G}_0(0)/4$, and 
$$t_{{\rm vg},0} = 1 - \Delta^{(R)}_0 ({\cal G}_0(0))^2 \eqno(2.8)$$
measures a distance from a {\it mean field} VG transition point 
$B_{{\rm vg},0}$. The quantity ${\cal G}_0(0)$ is selfconsistently determined 
by substituting eq.(2.7) into eq.(2.5) with $\omega=0$. 
Using it again, $t_{{\rm vg},0}$ valid near $B_{{\rm vg},0}$ is obtained 
(see eq.(2.12) below). 

Note that, in eqs.(2.7) and (2.8), the pinning strength $U_p$ (or 
$\Delta^{(R)}_0$) and the Matsubara frequency $\omega$ appear only as the 
combinations $x_p \equiv U_p \, ({\cal G}_0(0))^2/(2 \pi r_B^2)$ 
and $\gamma^{(R)} \omega \equiv \gamma \, {\cal G}_0(0) \omega$, 
respectively, and that, far below the quantum-thermal crossover line on the LLL fluctuation, defined by\cite{RI3,RI5} 
$$T \simeq T_{cr} = {{U_4} \over {2 \pi r_B^2 \gamma^2}} \eqno(2.9)$$
(see sec.2 in ref.14), $U_4$ appears only as the 
combination $x_4 \equiv U_4 {\cal G}_0(0)/(2 \pi^2 r_B^2 \gamma^2)$, as seen  
in $\sigma_{\rm vg}$. 
As a rule of Feynman diagram analysis in LLL,\cite{RI3} 
this is quite general and independent of the approximation used above. In fact, the relation $x_p \propto ({\cal G}_0(0))^2$, which is the same form as that 
occuring in the thermal 2D LLL case, arises because 
the pinning strength does not carry nonzero frequency, 
while the relation $x_4 \propto 
{\cal G}_0(0)$ is due to the fact\cite{RI3} that the purely dissipative 
quantum 
fluctuation raises the dimensionality of fluctuation by two and behaves like 
the thermal 4D LLL fluctuation. Based on this general rule, it is easy to 
generalize the above expressions of the lowest order in $U_p$ to the case 
with arbitrary pinning strength. To do this, we first note the 
definition of dynamical VG susceptibility\cite{RI1,RI4} 
$$\chi_{\rm vg}({\bf k}; \omega_1, \omega_2) = N_v^{-1} \int_{{\bf r}, {\bf R}} e^{{\rm i}{\bf k}\cdot{\bf R}} \, {\overline {< \Psi_{\omega_1}({\bf r}) \, \Psi^*_{\omega_1}({\bf r}+{\bf R}) > < \Psi_{\omega_2}({\bf r}+{\bf R}) \, \Psi^*_{\omega_2}({\bf r}) >}}$$ $$=(\beta N_v)^{-1} e^{-{\bf k}^2/2} \sum_{p, p'} e^{{\rm i}(p-p')k_x} {\overline {<\varphi_0(p, \omega_1) \, \varphi^*_0(p', \omega_1)> < \varphi_0(p'+k_y, \omega_2) \, \varphi^*_0(p+k_y, \omega_2)>}}, 
\eqno(2.10)$$ 
which appears in the expression of $G_{\rm vg}$, where the overbar denotes the 
random average. By this definition, 
the irreducible vertex in a diagrammatic representation 
of $\chi_{\rm vg}$ is found to carry a quantity 
$$\Delta^{(R)} = {{U_p} \over {2 \pi r_B^2}} \Delta(x_p; x_4) \eqno(2.11)$$
in the quantum regime $T < T_{cr}$ and consequently, the quantity 
$t_{{\rm vg},0}$ is given by replacing $\Delta^{(R)}_0$ in eq.(2.8) 
by $\Delta^{(R)}$, where $\Delta(x_p;x_4)$ is a unknown function of $x_p$ 
and $x_4$. Using, for simplicity, eq.(2.5) (with replacement $\Delta_0^{(R)} 
\to \Delta^{(R)}$), $t_{{\rm vg},0}$ near $B_{{\rm vg},0}$ and in $T \to 0$ 
limit is independent of the details of $\Delta^{(R)}$ and expressed in the 
form 
$$t_{{\rm vg},0} = {{\pi^2 \gamma r_B^2} \over {U_4}} \, {{B-B_{{\rm vg},0}} 
\over {B_{{\rm vg},0}}} \eqno(2.12)$$ 
up to O($B-B_{{\rm vg},0}$). The $B_{{\rm vg},0}$-value was examined in 
details elsewhere.\cite{RI5} 

In real disordered films, the critical fluctuation accompanying the second 
order VG transition at a critical field in $T \to 0$ limit 
is expected to be strong 
enough to change a temperature dependence near $B_{{\rm vg},0}$. 
To find how behaviors near $B_{{\rm vg},0}$ and the $B_{\rm vg}$-value itself 
are affected by the VG critical fluctuation, let us consider an effective 
action for the VG fluctuation field, as in the context of spin-glass, 
expressed as a tensor $Q_{\alpha, \alpha'}(\omega, \omega')$ with 
$\alpha \neq \alpha'$. Here $Q_{\alpha, \alpha'}$ depends on two frequency 
variables,\cite{FGG} 
reflecting that the pinning vertex carrying $\Delta^{(R)}$ does 
not convey nonzero frequency. Although a detailed diagrammatic analysis is 
needed to construct the action consistently with the derivation of eq.(2.12), 
this procedure can be bypassed here. First, it will be easily seen that, when 
performing a Landau expansion of the effective action with respect 
to $Q_{\alpha, \alpha'}$, a $Q^n$-order term is accompanied by $n$ 
frequency-summations. Further, according to the diagrammatic rules mentioned 
above, $\Delta^{(R)}$ appears everywhere as the combination $x^{(R)}_p \equiv 
\Delta^{(R)} ({\cal G}_0(0))^2$, which is a constant at $B_{{\rm vg}, 0}$ and 
in $T \to 0$ limit according to eq.(2.8) with $\Delta^{(R)}_0$ replaced 
by $\Delta^{(R)}$ and a modification to be done below. Then, the VG effective 
action should take the form 
$$S_{\rm eff} (Q) = \int d^2{\bf r} \biggl[ \sum_{\omega_1, \omega_2} 
\sum_{\alpha_1, \alpha_2} Q_{\alpha_2, \alpha_1}(\omega_2, \omega_1) \biggl( \,  t_{{\rm vg},0} + x^{(R)}_p (-\nabla^2) $$
$$+ g(\gamma^{(R)} |\omega_1|, \gamma^{(R)} |\omega_2|; \, x^{(R)}_p) \, \biggr) Q_{\alpha_1, \alpha_2}(\omega_1, \omega_2) + \sum_{m \geq 3} c_m 
(x^{(R)}_p)^{m/2} \prod_{j=1}^m \sum_{\omega_j} \sum_{\alpha_j} Q_{\alpha_j, \alpha_{j+1}}(\omega_j, \omega_{j+1}) \biggr], \eqno(2.13)$$
where the index $j=m+1$ implies $j=1$, $c_m$'s are constants, and the 
function $g$ satisfies $g(0, 0; x^{(R)}_p)=0$. First, it will be easily seen 
by considering this action at $T=0$ that any renormalization of the Gaussian 
mass $t_{{\rm vg},0}$ due to the interaction ($m \geq 3$) terms is 
accompanied only by the parameter $x^{(R)}_p$ and hence that the resulting renormalized one $t_{\rm vg}$ is given by 
$$t_{\rm vg} \equiv  t_{{\rm vg},0} + c_{\rm fl} = 1+c_{\rm fl} - x_p^{(R)} 
\simeq {{\pi^2 \gamma r_B^2} \over {U_4}} {{B-B_{\rm vg}} \over {B_{{\rm vg}, 0}}}, \eqno(2.14)$$ 
where $B_{\rm vg} \simeq B_{{\rm vg},0} (1 - c_{\rm fl} 
U_4/(\pi^2 r_B^2 \gamma))$ 
is the {\it fluctuation-corrected} VG transition field. As usual, a nonzero coefficient $c_{fl}$ is estimated in 
terms of a UV-divergent self energy term which, at the one loop level 
in eq.(2.13), occurs from a positive Hartree-like contribution of the $m=4$ 
term. Consequently, as expected, $c_{fl}$ becomes positive and 
thus, $B_{\rm vg}$ decreases with increasing the quantum fluctuation 
strength $U_4/(r_B^2 \gamma)$ according to the above-mentioned expression. 
Next, let us consider the action at nonzero $T$ ($< T_{cr}$) 
near $B_{\rm vg}$ where $|t_{\rm vg}| \ll 1$. This situation is {\it formally} 
similar to that in 3D systems 
at nonzero (but low) fields near $T_{c0}$ in which the 
so-called thermal XY scaling\cite{FFH,XY} $T-T_{c0} \sim B^{3/4}$ is expected below a crossover field (a line defined from the relation 
$\delta t_s=\delta t_d$ in ref.21) corresponding to $T_{cr}$ in the present 
issue. In the case, the field or the vortex density measures 
the inverse-square of the 
microscopic length $r_B$, and hence a scaling relation given above is expected 
by comparing the two lengths, $r_B$ and the correlation length of 
the pair-field $\sim (T-T_{c0})^{-\nu_{XY}}$ with $\nu_{XY} \simeq 2/3$, with 
each other. Similarly, in the present issue near $B_{\rm vg}$, 
we have only two variables $t_{\rm vg}(B)$ and $\gamma^{(R)}/\beta 
\propto \gamma/(\beta \sqrt{\Delta^{(R)}})$, and hence, the only 
$T$-dependent scaling variable is, in the present case, 
$(\xi_{\rm vg}(B, T \to 0))^z \gamma^{(R)}/\beta$, or equivalently, 
$X_{\rm vg} \equiv t_{\rm vg}(B) (\beta/\gamma^{(R)})^{1/z\nu}$, where $z$ is 
the dynamical exponent of the quantum 2D VG transition, and 
$\xi_{\rm vg}(B, T \to 0) = (t_{\rm vg}(B))^{-\nu}$ with an exponent 
$\nu > 0$ is the VG correlation length in $T \to 0$ limit written in unit 
of $r_B$. Thus, the scaling relation  
$$|B-B_{\rm vg}| \sim B_{{\rm vg}, 0} {{U_4} \over {\pi^2 \gamma r_B^2}} 
\biggl( {\gamma \over {\sqrt{\Delta^{(R)}}}} \, T \biggr)^{1/z \nu},  
\eqno(2.15)$$ 
measuring the field range of {\it quantum VG critical regime} 
at nonzero $T$, is expected near $B_{\rm vg}$, 
and the correlation length $\xi_{\rm vg}$ expressed in unit of $r_B$ should 
have, at nonzero $T$, the scaling form 
$\xi_{\rm vg}(B, T) = (t_{\rm vg}(B))^{-\nu} S_\xi(X_{\rm vg})$, where 
the scaling function $S_\xi$ has the limiting behaviors; $S_\xi(X_{\rm vg} \to 0) \to (X_{\rm vg})^\nu$, $S_\xi(X_{\rm vg} \to +\infty)$ approaches a positive 
constant. For convenience, we will assume $S_\xi(X_{\rm vg}) = (X_{\rm vg}/(c_\xi + X_{\rm vg}))^\nu$ leading to the correlation length 
$$\xi_{\rm vg}(B,T) = \biggl( t_{\rm vg}(B) + c_\xi 
\biggl( {\gamma \over {\sqrt{\Delta^{(R)}}}} \, T \biggr)^{1/z\nu} 
\, \biggr)^{-\nu} \eqno(2.16)$$ 
and yielding the correct limiting behaviors, 
where $c_\xi$ is a positive constant depending only on $x_p^{(R)}$ near $B_{\rm vg}$. Note that eq.(2.16) implies 
$\xi_{\rm vg}(B=B_{\rm vg}, T) \sim T^{-1/z}$. Further, although a line at finite $T$, on which $\xi_{\rm vg}$ apparently diverges, is defined in $B < B_{\rm vg}$ as a consequence of the scaling (2.15), the appearance of such a line is not peculiar to the present issue: For instance, although the thermal 
3D XY scaling in clean limit suggests a line at nonzero $B$ 
obeying this scaling, not a continuous transition but rather a first order 
(vortex lattice melting) transition occurs there. Similarly, the presence of 
such a line in the present 2D dirty case near $T=0$ 
does not contradict the absence\cite{MPAF} of a true 2D VG 
transition at $T>0$. Rather, eq.(2.16) itself will fail 
with decreasing $B$ away from the quantum VG critical regime, since 
the absence of the 2D VG 
transition at $T>0$ is usually guaranteed by a nonperturbative origin in a 
vortex solid.\cite{FGL}  

Now, let us examine the VG fluctuation contribution $G_{\rm vg}$ to the total 
conductance $G$. Near or below $H_{c2}(T)$-line at nonzero $T$, 
$G$ is generally expressed in the form\cite{RI7,RI2} 
$$G = G_n + G_s = G_n + \delta G_s + G_{\rm fl} + G_{\rm vg}, \eqno(2.17)$$
where $G_n$ and $G_s$ are, respectively, the quasiparticle contribution and the superconducting part of $G$, and $G_s$ consists of the Aslamasov-Larkin (AL) 
fluctuation term $G_{fl} + G_{\rm vg}$ and of other fluctuation 
contributions $\delta G_s$ excluded from the GL description. 
The AL contribution is further divided into 
the part $G_{\rm vg}$ due to VG fluctuation and the remaining one $G_{fl}$, 
which in clean limit becomes the vortex flow conductivity\cite{RI7} 
deep in the 
vortex liquid region at nonzero $T$. However, $G_{fl}$ in $B > B_{\rm vg}$ 
must vanish\cite{RI3} in $T \to 0$ limit, although this vanishing may not be 
detected at low enough $B$ in systems with weaker quantum fluctuation. 
Nonvanishing contribution in $T \to 0$ limit may arise from $G_n$ 
and\cite{NMR} $\delta G_s$, although discussing a {\it rigorous} $T=0$ 
result is beyond the scope 
of this paper. In this section, we focus primarily on $G_{\rm vg}$. Other 
fluctuation term $\delta G_s$ will be commented on in $\S 3$. 

To study $G_{\rm vg}$ (and $G_{fl}$) under a uniform current, we need the 
spatially averaged supercurrent $<{\bf j}_\Omega({\bf r})>_{\bf r}$ 
where $\Omega$ is the external frequency. In general, 
$<{\bf j}_\Omega({\bf r})>_{\bf r}$ will take the form 
$$<{\bf j}_\Omega({\bf r}_1)>_{{\bf r}_1} = {{2 \pi l^2} \over {\phi_0}} 
\beta^{-1} \sum_\omega < \, {\overline C}(|\omega|, |\omega + \Omega|; 
{\bf Q}_j) \, ({\bf Q}_1 + {\bf Q}^*_2) {\tilde \Psi}^*_\omega({\bf r}_2) 
{\tilde \Psi}_{\omega+\Omega}({\bf r}_1)|_{{\bf r}_1={\bf r}_2} 
\, >_{{\bf r}_1}$$
$$= {{2 \sqrt{2} \pi l^2 \beta^{-1}} \over {\phi_0 \, r_B}} \sum_\omega 
C(|\omega|, |\omega+\Omega|) \sum_p ( \, \varphi_1^*(p, \omega) \varphi_0(p, \omega+\Omega) ({\hat x}+{\rm i}{\hat y}) + \varphi_0^*(p, \omega) \varphi_1(p, \omega+\Omega) ({\hat x}-{\rm i}{\hat y}) \, ), \eqno(2.18)$$
where ${\bf Q}_j = -{\rm i}\partial/\partial {\bf r}_j + 2 \pi {\bf A}({\bf r}_j)/\phi_0$, $\varphi_1(p, \omega)$ a fluctuation field in the next lowest 
Landau level (NLL), and ${\tilde \Psi}$ the full pair-field prior to 
decomposing into the Landau levels. The $\omega$ and $\Omega$ dependences 
in $C(|\omega|, |\omega+\Omega|)$ are negligible when considering $G_{\rm vg}$ 
(and $G_{fl}$). 
Further, in obtaining the second line of eq.(2.18), terms consisting only of 
higher Landau levels were neglected by assuming a high field 
approximation\cite{RI1,RI2,RI4} in which a static vortex state is described 
within LLL. In particular, it is important to note that 
the NLL modes cannot participate in describing any 
static vortex solid for a reason of symmetry.\cite{RI6} Since 
this massive NLL modes inevitably 
appear in considering a response to a uniform 
current, some terms associated with NLL modes have to be added in the 
GL action, which are expressed by 
$$\delta S^n_p = \sum_\alpha \sum_\omega ({\cal G}_1(\omega))^{-1} 
|\varphi_1^{(\alpha)}(p, \omega)|^2 - {{U_p} \over {4 \pi r_B^2 N_v}} \sum_{\bf k} \sum_{\alpha, \beta} \biggl[ {{k^2} \over 2} f_{11}(k^2) \rho_1^{(\alpha)}
({\bf k}, 0) \rho_1^{(\beta) \, *}({\bf k}, 0)$$ 
$$+ \biggl( {{k_-} \over {\sqrt{2}}} 
f_{01}(k^2) \rho_1^{(\alpha)}({\bf k}, 0) \rho^{(\beta) \, *}({\bf k}, 0) 
+ {\rm c.c.} \biggr) \biggr], \eqno(2.19)$$
where $k_-=k_y+{\rm i}k_x$, 
and 
$$\rho_1^{(\alpha)}({\bf k}, 0) = \sum_{\omega, p} e^{{\rm i} p k_x - {\bf k}^2/4} \varphi_1^{(\alpha) \, *}(p-k_y/2, \omega) \, 
\varphi_0^{(\alpha)}(p+k_y/2, \omega). \eqno(2.20)$$ 
For convenience of presentation, the NLL 
propagator ${\cal G}_1(\omega) = <|\varphi_1^{(\alpha)}(p, \omega)|^2> = (\mu_1(0) + \gamma^{(1)}|\omega|)^{-1}$ is assumed here to have been renormalized in a proper way. The bare mass $\mu_1(0)$ is of order unity near $H_{c2}(0)$. 

The terms $G_{fl} + G_{\rm vg}$ of conductance are obtained in terms of Kubo formula\cite{RI3} 
$$\biggl({{\pi r_B^2} \over {M_1}} \biggr)^2 R_Q  (G_{fl} 
+  G_{\rm vg}) $$ $$= \biggl(- {\partial 
\over {\partial |\Omega|}} \biggr) (\beta N_v)^{-1}  \! \sum_{p, p', \omega} {{\overline {< \varphi_0(p, \omega) \varphi^*_0(p', \omega) \varphi_1(p', \omega+\Omega) \varphi^*_1(p, \omega+\Omega)>}} \over {({\cal G}_1(0))^2}}  
\biggl|_{\Omega \to 0},  \eqno(2.21)$$
where $R_Q=\pi \hbar/2 e^2$ is the resistance quantum. 
In the present high field approximation, ${\cal G}_1(0)$ 
appears in both $G_{\rm vg}$ and $G_{fl}$ only as the combination 
$M_1 \equiv 2 \pi l^2 C(0, 0) {\cal G}_1(0)$. Below, it will be shown that, 
deep in the vortex liquid regime, this factor takes the universal value 
$$M_1 = \pi r_B^2 \eqno(2.22)$$
independent of microscopic details as a consequence of gauge-invariance. Deep 
in the vortex liquid regime, the AL conductance $G_{fl}$ in the pinning-free 
case can be assumed to be given by the vortex flow expression.\cite{RI7} 
Since the 
vortex flow in a {\it pinning-free} system is not affected by the 
vortex-solidification at $T_m(B)$, we can focus on the mean field result of 
the vortex flow conductance, which is most easily derived in terms of a 
harmonic action\cite{RI8} 
$$\delta S_{\rm har} = {{\beta <|\Psi_{\rm MF}({\bf r})|^2>_{\bf r}} 
\over {2 r_B^2}} \sum_\Omega \biggl( \gamma|\Omega| |{\bf s}_L|^2 
+ ({\cal G}_1(0))^{-1} \biggl|{\bf s}_L + {{2 \pi} \over {\phi_0}} 
(\delta{\bf A} \times {\hat z})r_B^2 \biggr|^2 \biggr), \eqno(2.23)$$
where ${\bf s}_L$ denotes a {\it uniform} displacement of vortices, 
and $\Psi_{\rm MF}$ is the mean field solution of vortex solid. The combination $r_B^{-2}({\hat z} \times {\bf s}_L) + (2 \pi/\phi_0) \delta{\bf A}$ in 
eq.(2.23) is a consequence of the gauge-invariance and ensures the Josephson relation ${\bf E} = -\partial \delta{\bf A}/\partial t = - {\bf v}_L \times {\bf B}$, where ${\bf v}_L=\partial {\bf s}_L/\partial t$ is the vortex velocity. By 
substituting $B {\bf s}_L = {\hat z} \times \delta{\bf A}$ minimizing the above second term into the first term, the vortex flow conductivity is obtained as the coefficient of thus obtained first term, which is clearly independent 
of ${\cal G}_1(0)$. On the other hand, using eq.(2.21) (with no pinning disorder effect) in the mean field approximation where $\varphi_0$'s $\omega$-dependence is absent, one obtains a 
conductance expression $\propto M_1^2$. Under the condition that these two 
expressions are identical with each other, eq.(2.22) follows. In fact, using 
the general fact,\cite{RI8} already used in writing down eq.(2.23), that 
the {\it uniform} fluctuation belonging to NLL around the 
mean field solution $\Psi_{\rm MF}$ is nothing but a uniform displacement 
solution with amplitude $s_{L, y}-{\rm i}s_{L,x}$, the averaged supercurrent 
(2.18) is found to be 
proportional to $2\pi l^2 C(0,0) ({\bf B} \times {\bf s}_L + 
\delta{\bf A})$ up to O(${\bf s}_L$) and O($\delta{\bf A}$). By identifying 
this with that following from eq.(2.23), we again obtain eq.(2.22). Although, due to a pinning disorder effect on the static NLL mode, the pinning-free result 
(2.22) may be subject to a subtle change, we argue in terms of eq.(2.8) 
that, as far as ${\cal G}_1(0) \ll {\cal G}_0(0)$ is safely satisfied, 
eq.(2.22) is quantitatively valid near $B_{\rm vg}$ in the present high field 
case. 

Before examining $G_{\rm vg}$ further, a form of $\chi_{\rm vg}$ needs to be 
determined. In the Gaussian (or mean field) approximation, it is written as 
$$\chi_{{\rm vg}, 0} ({\bf k}; \omega_1, \omega_2) = ({\cal G}_0(0))^2 \, 
(\xi_{{\rm vg},0})^2  $$ $$\times \biggl(1 + {{({\bf k} \xi_{{\rm vg},0})^2} 
\over 2} + {{2 (\xi_{{\rm vg},0})^4 \gamma^{(R)}|\omega_1|} \over {1 + \sqrt{1 + 4(\xi_{{\rm vg},0})^4 \gamma^{(R)}|\omega_1|}}} + {{2 (\xi_{{\rm vg},0})^4 
\gamma^{(R)} |\omega_2|} \over {1 + \sqrt{1+4(\xi_{{\rm vg},0})^4 \gamma^{(R)} 
|\omega_2|}}} \biggr)^{-1}, \eqno(2.24)$$
where $\xi_{{\rm vg}, 0} = t_{{\rm vg},0}^{-1/2}$ is the mean field 
VG correlation length in unit of $r_B$, and eq.(2.7) was used. 
To go beyond the mean field analysis for studying $G_{\rm vg}$ and enter 
the quantum VG {\it critical} regime present in $T < T_{cr}$, we invoke the ordinary scaling hypothesis\cite{FFH,RI1,RI9} 
for the VG correlation function on the basis of the above mean 
field expression as follows: 
$$\chi_{\rm vg}({\bf k}; \omega_1, \omega_2) = c_g \, (\xi_{\rm vg} \, 
{\cal G}_0(0))^2 \, S_\chi({\bf k} \xi_{\rm vg}; \gamma^{(R)} (\xi_{\rm vg})^z |\omega_1|, \gamma^{(R)} (\xi_{\rm vg})^z |\omega_2|), \eqno(2.25)$$
where the correlation length $\xi_{\rm vg}$ is given by eq.(2.16), and the 
scaling function $S_\chi$ and a positive coefficient $c_g$, as well as $c_\xi$ in eq.(2.16), may depend 
on $x^{(R)}_p$ according to the action (2.13). 

Now, the terms corresponding to $G_{\rm vg}$ in r.h.s. of eq.(2.21) will be 
examined. First, closely following the analysis\cite{RI9,RI1} used for the thermal glass transitions, let us consider 
the contributions $G_{\rm vg}^{(1a)}$ of the diagram Fig.1 (a) 
and $G_{\rm vg}^{(1b)}$ of a sum of the family of Fig.1 (b) to eq.(2.21) 
with the coefficient (2.22), which are expressed as 
$$R_Q G_{\rm vg}^{(1a)} = {{U_p} \over {2 r_B^2}} \int_{\bf k} k^2 e^{-k^2/2} f_{11}(k^2) \biggl(- {\partial \over {\partial |\Omega|}} \biggr) \beta^{-1} \sum_\omega \chi_{\rm vg}({\bf k}, \omega, \omega+\Omega)$$ $$= \pi x_p \, c_g(x_p^{(R)}) \, f_{11}(0) \, \xi_{\rm vg}^{-2} \, (\gamma^{(R)} \xi_{\rm vg}^z 
\beta^{-1}) \int_{\bf k} k^2 \sum_n S'_\chi ({\bf k}; \, 2 \pi |n| (\gamma^{(R)} \xi_{\rm vg}^z \beta^{-1}) \, ), \eqno(2.26)$$

$$R_Q G_{\rm vg}^{(1b)} = {{U_p^2} \over {4 \pi r_B^4}} c_{01} 
\biggl(-{\partial \over {\partial |\Omega|}} \biggr) \beta^{-1} \sum_\omega ({\cal G}_0(\omega) + {\cal G}_0(\omega+\Omega))^2 \int_{\bf k} \chi_{\rm vg}({\bf k}; \omega, \omega+\Omega)$$  $$= 4 \pi c_{01} \, c_g(x_p^{(R)}) \,x_p^2 \, 
(\gamma^{(R)} \xi_{\rm vg}^z \beta^{-1}) \sum_n \int_{\bf k} 
S'_\chi({\bf k}; \, 2\pi |n| \gamma^{(R)} \xi_{\rm vg}^z \beta^{-1}), 
\eqno(2.27)$$ 
where 
$c_{01}=\int_{\bf k} k^2 (f_{00}(k^2) f_{11}(k^2) - (f_{01}(k^2))^2) \, e^{-k^2/2}$, and $\sum_\omega S'_\chi({\bf k}; \, |\omega|) 
= -(\partial/\partial|\Omega|)$ $\sum_\omega S_\chi({\bf k}; \, |\omega|, |\omega + \Omega|)|_{\Omega=0}$. 
Although we have assumed in eqs.(2.26) and (2.27) 
the "pinning lines" unrelated to the 
VG susceptibility to carry not $\Delta^{(R)}$ but $U_p/2 \pi r_B^2$, 
this simplification does not affect the conclusion given below, that $R_Q G_{\rm vg}^{(1b)}$ in the dirty limit takes a universal value independent of $T$ 
at $B_{\rm vg}$, because of the relations (2.11) and (2.34) (see below). 

So far, no microscopic (electronic) model leading to the dissipative GL 
action has been specified, and a $T=0$ limit of the resulting GL action has 
been simply assumed. Although we will consider below a simplified BCS 
hamiltonian with a short-ranged repulsion and a nonmagnetic potential disorder 
in order to see how a $T$-insensitive GL action can be realized, it is 
instructive to 
first start from the dirty limit with no electron-repulsion. 
In this case, the expressions of coefficients $U_4$, $\gamma$, $\mu(0)$, $\mu_1(0)$, and $C(0,0)$ are available in the literatures\cite{RI3,Parks} 
and given by 
$$U_4 = 8 \pi \tau^3 (\beta N(0))^{-1} \sum_{\epsilon > 0} (\Gamma(2\epsilon; B))^3, \eqno(2.28)$$
$$\gamma=4 \pi \tau^2 \beta^{-1} \sum_{\epsilon > 0} (\Gamma(2\epsilon; B))^2, \eqno(2.29)$$
$$\mu(0) = {\rm ln}\biggl({T \over {T_{c0}}}\biggr) + 4 \pi \tau \beta^{-1} 
\sum_{\epsilon > 0} (\Gamma(2\epsilon; 0) - \Gamma(2\epsilon; B)), \eqno(2.30)$$ $$\mu_1(0) = \mu(0) + 4 \pi \tau \beta^{-1} \sum_{\epsilon > 0} (\Gamma(2\epsilon; B) - \Gamma(2\epsilon; 3B)), \eqno(2.31)$$ 
and $$C(0,0) = 2 \pi \tau \beta^{-1} \sum_{\epsilon > 0} \Gamma(2\epsilon; B) \Gamma(2\epsilon; 3B), \eqno(2.32)$$
where $\Gamma(2\epsilon;B) = (2|\epsilon|\tau + \pi l^2 B/\phi_0)^{-1}$, and $\epsilon$ denotes a Matsubara frequency for fermions. Note that the relation (2.22) is satisfied just on the $H_{c2}(T)$ line where $\mu(0)=0$, implying that, 
in the dirty limit, the renormalized $\mu_1(0)$ should approach $4 \pi \tau \beta^{-1} \sum_{\epsilon > 0} (\Gamma(2\epsilon; B) - \Gamma(2\epsilon; 3B))$ with decreasing $B$. In $T < T_{cr}^{\rm mf}$, the $T$-dependence of $\Gamma(2\epsilon;B)$ is cut off by the $B$-dependence, and the above coefficients become insensitive to $T$ in the manner $U_4 \to 4 \tau^2 r_B^4/(N(0) \, l^4)$, $\gamma \to 2 \tau (r_B/l)^2 \equiv \gamma^{(0)}(T=0)$, $\mu(0) \to {\rm ln}(B/H_{c2}(0))$, and $\mu_1(0) \to {\rm ln}(3B/H_{c2}(0))$. On the other hand, the pinning strength $U_p$ is known only in $T < T_{cr}^{\rm mf}$ and given by\cite{RI5} 
$$U_p \simeq r_B^2 \biggl({\tau \over {N(0) l^2}} \biggr)^2. \eqno(2.33)$$
In terms of these $T$-independent GL coefficients, a "$T=0$" critical field 
$B_{\rm vg}$ can become well-defined within the dirty limit. Further, at lower temperatures than $T_{cr}$ which is estimated in dirty limit 
as $\simeq 4 \pi T_{cr}^{\rm mf}/(E_{\rm F} \tau)$ ($\propto B$), 
we have the relation 
$$x_p={{\pi^3} \over 2} x_4^2. \eqno(2.34)$$ 
As a result of eq.(2.34), $x_p^{(R)}$ depends only on $x_p$. 
Just at $B_{\rm vg}$, the combination $\gamma^{(R)} (\xi_{\rm vg})^z/\beta$, 
as well as $x_p^{(R)}$ and $x_p$, becomes a constant independent of $T$ and of 
material parameters according to eqs.(2.11) and (2.14). 
Therefore, $G_{\rm vg}^{(1b)}(B=B_{\rm vg})$ 
is a {\it universal} constant divided by 
$R_Q$ in $T < {\rm Min}(T_{cr}, \, T_{cr}^{\rm mf})$. 
Similarly, $R_Q G_{\rm vg}^{(1a)}(B=B_{\rm vg})$ 
becomes $(\gamma T/\sqrt{\Delta^{(R)}})^{2/z}$ 
multiplied by a universal positive constant. 

It should be mentioned that we cannot verify directly whether the (universal) 
constant $c_{01}$ is positive or not, because the functional forms of 
$f_{00}$, $f_{11}$, and $f_{01}$ are not completely known, although, by 
definition, the VG contribution $G_{\rm vg}$ to the conductance {\it must} be 
positive. On the other 
hand, we have a unlimitedly large number of diagrams contributing to 
$R_Q G_{\rm vg}(B=B_{\rm vg})$ in the same way as Fig.1(b). 
All of them can be seen as such diagrams\cite{RI9} that, 
according to the already-mentioned LLL diagrammatic rule, the vertex 
correction {\it unrelated} to the VG susceptibility is of 
higher order in $x_p$ and $x_4$ (in $T < T_{cr}$) compared with those 
in Fig.1. In this sense, Fig.1(a) is the lowest order term, and the diagram in 
Fig.1(b) is the next lowest order term in $x_p$. 
However, {\it just at} $B=B_{\rm vg}$ where $x_p$ and $x_4$ take constant 
values, the diagram of Fig.1(a) becomes of the same order as that of 
Fig.1(b) except the extra power in $\xi_{\rm vg}^{-2}$. 
The same thing holds in the (formally) higher order diagrams in $x_p$ 
and $x_4$, and hence, they also contribute to a universal value of 
$R_Q G_{\rm vg}(B=B_{\rm vg})$ together with a sum of the family of Fig.1(b). 
Since, unfortunately, we have no resummation scheme, useful at the critical 
point, 
for judging which of those diagrams should be adopted or may be neglected, it 
is difficult, as in the argument\cite{MPAF,FGG} on a $T=0$ critical conductance $R_Q G_s(T=0, \omega \to 0)$, to estimate here a concrete value of 
$R_Q G_{\rm vg}(B=B_{\rm vg})$. We 
can just conclude that, in the present dirty limit, $R_Q G_{\rm vg}(T \to 0$, 
$B=B_{\rm vg})$ is a universal positive number. 

Of course, a universal critical $G_{\rm vg}$ obtained above is not a 
consequence of the $T=0$ scaling argument.\cite{MPAF,FGG} Actually, the 
$\omega$-summation in eq.(2.27) was not changed above into a frequency-integral because $\gamma^{(R)} \xi_{\rm vg}^z \beta^{-1}$ is finite in $T \to 0$ limit 
and at $B_{\rm vg}$, implying that $G_{\rm vg}^{(1b)}(B=B_{\rm vg})$ is not a 
dc limit of a $T=0$ conductance\cite{MPAF,FGG} but a dc conductance in the 
quantum regime $0< T < T_{cr}$. 

However, it will be difficult to explain available resistivity data in terms 
only of the results in the dirty limit with no electron-repulsion. 
As examined in ref.19, the mean field value $B_{{\rm vg},0}$ calculated in the 
dirty limit 
seems to increase with increasing $(E_{\rm F}\tau)^{-1}$ due to the 
corresponding increase of $H_{c2}^d(0)$, while the data in ref.4 have shown a 
trend opposite to this. Although the resulting fluctuation-corrected field 
$B_{\rm vg} \sim B_{{\rm vg},0}(1 - 2 c_{\rm fl}/(\pi E_{\rm F} \tau) \,)$ 
(see the sentence below eq.(2.14)) may decrease with 
increasing $(E_{\rm F}\tau)^{-1}$ depending on $E_{\rm F}\tau$-values, 
it will be difficult to understand a dependence\cite{Gold} of $B_{\rm vg}$ 
on the film thickness $d \sim k_{\rm F}^{-1} R_Q/(R_r E_{\rm F} \tau)$ 
in the dirty limit with no 
electron-repulsion, where $R_r$ is the high temperature sheet resistance. Once 
the interplay between the electron-repulsion and disorder is taken into 
account, however, $H_{c2}(0)$ and hence, $B_{\rm vg}$ decrease\cite{RI5} 
with increasing the electron-repulsion strength\cite{Fin} $\lambda_1 \simeq R_r/(8 \pi R_Q)$. Further, a nonzero $\lambda_1$ results in a failure of the equality (2.34), and thus, the $G_{\rm vg}(T \to 0)$ value at a critical field is not 
universal any longer but will depend on $\lambda_1$. 

We argue that the properties in the close vicinity of $T=0$ (corresponding to 
the region below $T_{\rm rep}$) of disordered thin superconducting films 
have {\it not} been examined so far experimentally. 
According to the previous works\cite{RI5,RI10}, the interplay between an 
electron-repulsion and disorder appears in the GL action in two different ways: The GL coefficients $U_4$, $\mu(0)$, $\mu_1(0)$, $C(0,0)$, and $U_p$ are 
convergent in low $T$ limit at each order of the $\lambda_1$-perturbation 
series, and their $T$-dependences are controlled, through the denominator of Cooperons, by the $|\epsilon|$ (Matsubara frequency) value of the order 
$l^2/(8 \pi \tau r_B^2)$. Namely, their $T$-dependences are lost, as well as 
those in the dirty limit, below $T_{cr}^{\rm mf}$ independent 
of $\lambda_1$. In contrast, $\gamma$ at low enough $T$ 
is expanded\cite{RI5,RI10} in powers 
of $\lambda_1 {\rm ln}(T/T_{cr}^{\rm mf})$ because this quantity is dominated 
by the lowest $|\epsilon|$ values, and consequently, a $T$-dependence of 
$\gamma$ induced by the electron-repulsion will become remarkable rather 
near $T=0$ below\cite{RI5} $T_{\rm rep} \simeq T_{cr}^{\rm mf} 
\exp(-\lambda_1^{-1})$. Since, as far as we know, $R_r < R_Q$, 
or equivalently $8 \pi \lambda_1 < 1$, is satisfied 
in real thin films with an FSI 
behavior, $T_{\rm rep}$ will lie much below $T_{cr}^{\rm mf}$ and seems to be 
inaccessibly low in real systems. Then, it is reasonable 
to assume the FSI behavior seen in real experiments to be a phenomenon in 
the intermediate region $T_{\rm rep} < T < T_{cr}^{\rm mf}$, where all GL 
coefficients are insensitive to $T$ so that an {\it apparent} VG critical 
field\cite{RI5} $B_{\rm vg}^*$ and a nonuniversal constant $R_Q G_{\rm vg}(B=B_{\rm vg}^*)$ are well-defined. 

In $\S 4$ of ref.19, a computation result on $\gamma$ was given supporting the 
argument on the presence of the intermediate temperature region, and it was 
suggested that the temperature scale corresponding to $T_{\rm rep}$ will be 
below $0.1$ $T_{\rm cr}^{\rm mf}$ in the cases with realistic $R_r$-values. 
Unfortunately, it was difficult to judge 
whether $\gamma$ remains positive or vanishes in low $T$ limit, i.e., 
in $T < T_{\rm rep}$. We simply expect here that $\gamma$ will significantly 
decrease on cooling below $T_{\rm rep}$ 
and hence that, within the model of purely dissipative dynamics, 
the true $B_{\rm vg}$ will lie at a much lower field than $B_{\rm vg}^*$ 
(see Fig.3 below).\cite{RI5} What we wish to emphasize is that, as far as $8 
\pi \lambda_1 < 1$ is satisfied in real systems, the $T$-dependence of a 
microscopic origin is remarkable rather at extremely low temperatures 
below $T_{\rm rep}$, and hence that any attempt, such as ref.6, to explain 
the FSI behaviors by assuming a $T=0$ bosonic model is not justified. 

\section {Description of 2D Resistive Behavior near $B_{\rm vg}^*$}

In this section, we discuss the resistivity curves around $B_{\rm vg}^*$ and 
in $T < T_{cr}^{\rm mf}$ on the basis of the results in $\S 2$ and give some 
results relevant to comparing with experimental data. Again, the GL 
coefficients will be assumed for a moment to be insensitive 
to $T$ so that a critical field $B_{\rm vg}$ may be well-defined. 

First, let us start with the resistive behaviors far 
above $B_{\rm vg}$ ($B > B_{\rm vg}$) where the Gaussian approximation, 
illustrated as eq.(2.24), for the VG fluctuation may be used. 
Assuming the low frequency behavior 
to be essential even in the Gaussian region, we will keep, for simplicity, 
only O($|\omega_j|$) ($j=1$, $2$) terms in eq.(2.24), and 
this $\chi_{{\rm vg}, 0}$ will be substituted 
into the first line of eq.(2.27). By arranging the $\omega$-summation and 
performing the $|\Omega|$-derivative, we obtain
$${{R_Q G_{{\rm vg}, 0}^{(1b)}} \over {2 c_{01}}} = x_p^2 \, \beta^{-1} \biggl( \gamma^{(R)} \xi_{{\rm vg}, 0}^4 + {5 \over 2} \gamma^{(R)} \xi_{{\rm vg}, 0}^2 \, {\rm ln}(1 + c_c^{-2} \xi_{{\rm vg},0}^2) + ({\cal G}_0(0))^{-2} \sum_{\omega > 0} {\partial \over {\partial \omega}} \biggl[ - ({\cal G}_0(\omega))^2$$ $$\times {\rm ln}\biggl({{1 + c_c^{-2} \xi_{{\rm vg}, 0}^2 + 2 \omega \gamma^{(R)} \xi_{{\rm vg}, 0}^4} \over {1 + 2\omega\gamma^{(R)} \xi_{{\rm vg},0}^4}} \biggr) + {\rm ln}(1 + c_c^{-2} \xi_{{\rm vg}, 0}^2) {{{\cal G}_0(\omega)} \over 2} ({\cal G}_0(\omega) + {\cal G}_0(0)) \ \biggr] \biggr), \eqno(3.1)$$
which obviously vanishes in $T \to 0$ ($\beta^{-1}\gamma^{(R)} \xi_{{\rm vg},0}^4 \to 0$) limit. Similarly, one can verify that 
$G_{{\rm vg},0}^{(1a)}$ also vanishes at $T=0$ if taking account of the 
frequency dependence of ${\cal G}_1$. The additional 
${\rm ln} \xi_{{\rm vg},0}^2$ dependences arise from an {\it upper} 
cutoff $(c_c r_B)^{-1}$ (with a constant $c_c$ of order unity) of the 
$|{\bf k}|$-integral. Judging from a similar situation one encounters 
in deriving 2D AL fluctuation conductance in $T \to 0$ limit and at $B=0$, 
we believe that this divergence is specific to the present direct 
frequency-summation and may be avoided by the standard analytic continuation 
which we have not tried. However, this technical issue does not affect our 
conclusion that, as well as $G_{fl}$, $G_{\rm vg}(B > B_{\rm vg})$ vanishes 
in $T \to 0$ limit, because, as shown in ref.14, each term of 
perturbation series of fluctuation conductivity examined within a quantum GL action vanishes in $T \to 0$ irrespective of the presence or absence of a pinning-disorder term in the action. In $B \gg B_{\rm vg}$, such a 
perturbation series should become more convergent with approaching $T=0$, 
and hence, we can conclude that, as well as each term of the perturbation 
series, the resummation results of the perturbation series, i.e., $G_{fl}$ 
and $G_{\rm vg}$ themselves in $B > B_{\rm vg}$, also vanish 
in $T \to 0$ limit. 

On the other hand, in $B < B_{\rm vg}$ and out of the quantum critical regime 
defined by eq.(2.15), $\xi_{\rm vg}$ grows with decreasing $T$ or $B$ 
and hence, the VG fluctuation 
becomes "classical" even at low temperatures below $T_{cr}$ in which the pair-field fluctuation, with higher energy than the VG fluctuation, is 
of a quantum character. In fact, as eq.(3.1) suggests, one may expect the 2D 
classical behavior\cite{FFH} $G_{\rm vg} \sim \beta^{-1} \gamma^{(R)} \xi_{\rm vg}^z$ 
within the present analysis. However, the classical (i.e., thermal) 2D VG 
transition is washed out, e.g., by a nonperturbative effect such as the free vacancies or interstitials in the vortex solid.\cite{FGL} 
Therefore, the region in which the classical scaling behavior $G_{\rm vg} 
\sim \xi_{\rm vg}^z$ is visible may be quite narrow. 

According to eq.(2.27), the scaling behavior of $R_Q G_{\rm vg}$ 
$$R_Q G_{\rm vg} = {\cal U} \biggl( c_u \, {{B-B_{\rm vg}} \over {B_{{\rm vg}, 0}} } \biggl({{T_{c0}} \over T} \biggr)^{1/z\nu} \biggr) \eqno(3.2)$$
is expected in the quantum VG critical regime defined by the relation (2.15). 
Here, ${\cal U}(x=0)$ is a positive and nonuniversal constant according to the 
result in $\S 2$, and 
$$c_u = {{\sqrt{\Delta^{(R)}} r_B^2} \over {U_4 T_{c0}}} \biggl({{\gamma 
T_{c0}} \over {\sqrt{\Delta^{(R)}}}} \biggr)^{(z\nu-1)/z\nu}. 
\eqno(3.3)$$
Further, if describing the "Gaussian" region, discussed above, in terms of 
eq.(3.2), the limiting behaviors ${\cal U}(x \to +\infty) \to 0$ and 
${\cal U}(x \to -1) \to (x+1)^{- z \nu}$ have to be satisfied. Therefore, 
the resulting $(R_Q G_{\rm vg})^{-1}$ v.s. $T$ curves 
around $B_{\rm vg}$ in $B$-$T$ 
plane are, just like the original speculation\cite{Steve} 
based on some $B=0$ results, similar to that of a "renormalization-group 
flow" near a fixed point. Further, we note that, if $z\nu > 1$, 
the $\gamma$-dependence 
of $c_u$ will be opposite to a {\it naive} expectation on a strength of 
quantum critical fluctuation. In the case with a quantum normal-Meissner (i.e., $B=0$) transition, for instance, one would expect the width of the quantum 
critical 
regime at a fixed $T$ to scale like $\gamma^{1/z\nu}$ and hence to become 
narrower for a stronger quantum fluctuation. In contrast, in the present 
$B > 0$ problem, the quantum VG critical region becomes wider as 
the {\it pair-field} fluctuation is enhanced. Details of the low $T$ regions in $B$-$T$ phase diagram of thin films are sketched in Fig.2. 

As shown previously,\cite{RI3} $R_Q G_{fl}$ calculated in the 
{\it pinning-free} ($U_p=0$) case shows, around a field above $B_m(0)$,  
"fan-shaped" resistivity curves similar to but more moderate than that 
of $G_{\rm vg}$ near $B_{\rm vg}$. This follows from the facts that, 
when $\gamma$ and hence the melting line $B_m(T)$ 
in the quantum regime are insensitive to $T$, a "flat" $G_{fl}^{-1}$ curve of 
the order of $R_Q$ is realized at a field above $B_m$ and that, in 
higher fields, $G_{fl}^{-1}$ shows an insulating behavior reflecting 
$G_{fl}(T=0)=0$, while it yields, in lower fields closer to $B_m$, the classical vortex flow behavior which is insensitive to $T$ at such low 
temperatures (see, for instance, Fig.4 of ref.14). Although an inclusion of the vortex pinning effect will slightly change this behavior of $G_{fl}$ 
particularly close to $B_m$, it is generally 
questionable to interpret real resistance curves by neglecting the presence 
of $G_{fl}$ and identifying the resistance only with 
$G_{\rm vg}^{-1}$ because, as mentioned above, $G_{fl}$ can have a magnitude of the same order as $G_{\rm vg}$ near $B_m$. 
The fluctuation corrections to $G_n$, $\delta G_s$, will be taken into account 
later, and for a moment we identify the total conductance with the 
bosonic contribution $G_{fl} + G_{\rm vg}$ plus the normal 
contribution $G_n$. 

Now, we discuss typical resistance data under the assumptions 
that $T_{\rm rep}$ is inaccessibly low (see the end part of $\S 2$) and that an apparent critical field $B_{\rm vg}^*$, well-defined in $T > T_{\rm rep}$, 
will lie above $B_m$. Due to the former assumption, all the GL coefficients and hence, $B_{\rm vg}^*$ are insensitive to $T$ in $T_{\rm rep} < T 
< T_{cr}^{\rm mf}$. One will see soon that the latter assumption has already 
been verified\cite{Okuma2,Valles} in some data. First, let us start 
with the cases with weak disorder (small $R_r$), 
in which the quantum fluctuation is also weak 
simultaneously and $B_{\rm vg}^*$ will lie near but above $B_m$ (see below). 
In this case, the contributions of $G_n$ and/or $G_{fl}$ 
occupy a large weight of the total conductance near $B_{\rm vg}^*$, 
and there may be no clear indication of quantum correction\cite{HF} 
($\sim - \lambda_1 {\rm ln}(1/T\tau)$) 
to the familiar residual behavior (insensitive to $T$) of $G_n$ 
at accessible temperatures above $T_{\rm rep}$. Since $T_{cr}^{\rm mf} 
\tau < B l^2/\phi_0 < 1$ in fields of our interest, the negligible quantum correction to $G_n$ implies that the neglect of the region below $T_{\rm rep}$ is 
justified. Further, since 
$G_{fl}$ just above $B_m$ in the present case is, as mentioned in the last 
paragraph, insensitive to $T$, the background contribution $G_n+G_{fl}$ is 
expected to be independent of $T$ and to weakly depend on $B$. On the other 
hand, according to eq.(3.3), the quantum VG critical region at a 
fixed $T$ seems to be narrower at weaker disorder: As a rough estimation in this weak disorder case, if any $\lambda_1$-dependence is neglected in 
eq.(3.3), $c_u$ certainly decreases with increasing disorder as far as 
$z\nu > 1$, as seen in most of data. By taking account of these contributions to the conductance altogether, the FSI behavior (3.2) is expected to be 
visible in a narrow VG critical field range around $B_{\rm vg}^*$. 
We believe that this will be an appropriate explanation, for instance, 
to the data in MoSi films\cite{Okuma2} where the total resistance value 
near $B_{\rm vg}^*$ was remarkably suppressed due to a large weight 
of $G_{fl} + G_n$ insensitive to $T$. In fact, in ref.4, the field $B_0$ 
below which the thermal activation barrier\cite{FGL} $\sim {\rm ln}B^{-1}$, 
arising from the motions of vacancies and interstitials 
in a (short-range ordered) vortex solid, remains nonvanishing 
lies just below the critical field corresponding to $B_{\rm vg}^*$. Since $B_0$ should be essentially the same as $B_m$, it justifies the above assumption 
that $B_{\rm vg}^*$ lies {\it just} above $B_m$. 

It is possible that, in systems with still weaker disorder, even the quantum 
fluctuation behavior of $G_{fl}$ is not seen. In such a case with inaccessibly 
low $T_{cr}$, the flat behavior of $G_{\rm vg}$ should not be seen 
consistently. This situation corresponds to the data in ref.33 where the only quantum behavior was seen in a vortex flow behavior, which is itself the {\it classical} behavior of $G_{fl}$, but suggestive of a quantum tunnelling effect. 
Clearly, the metallic behavior below $H_{c2}$ there\cite{kasu} is {\it not} a 
reflection of a $T=0$ phase diagram: When the sample disorder is weaker, 
i.e., $T_{cr}$ is low enough, 
one needs to enter a lower temperature region to find quantum VG behaviors. 
Theoretically, an intermediate vortex liquid {\it phase} with finite 
resistance much smaller than $G_n^{-1}$ is absent\cite{RI3} 
in 2D {\it homogeneously} disordered films at $T=0$ (see also $\S 4$). 

Returning again to systems with visible quantum critical behaviors of $G_s$, 
in turn, let us consider situations with stronger disorder 
(larger $R_r$). In this case, a decrease of $G_{fl}$ and $G_n$ on cooling will 
affect the total conductance: The total conductance 
at $B_{\rm vg}^*$ will decrease on cooling until a constant limiting 
value $G \simeq G_{\rm vg}(B_{\rm vg}^*)$ is reached. Such a behavior that the 
resistance {\it just at} $B_{\rm vg}^*$ is not flat but increases on cooling 
until the lowest (accessible) temperature is reached has been observed in 
various materials.\cite{Gan1,Gold} If such data are explained according to 
this scenario, the vanishing contribution $G-G_{\rm vg}$ in $T \to 0$ will 
not be accompanied by the scaling behavior (2.15). 
However, an insulating behavior {\it at} $B_{\rm vg}^*$ may appear even 
if $G \simeq G_{\rm vg}$: As illustrated by eq.(2.26), a sub-leading term of $G_{\rm vg}(B_{\rm vg}^*)$ will behave like $\sim (\xi_{\rm vg}(B=B_{\rm vg}^*))^{-2} \propto T^{2/z}$. If the sum of such terms is positive just 
like eq.(2.26) itself, this also becomes an origin of an insulating behavior 
at $B_{\rm vg}^*$. We note that, in this case, the sub-leading terms in 
the quantum VG critical region are multiplied by a critical 
scaling function like ${\cal U}(x)$ in eq.(3.2) and hence that 
they can be discriminated from an insulating behavior of 
the noncritical term $G-G_{\rm vg}$ mentioned earlier. 

In ref.7, nonmonotonic resistance curves (insulating at intermediate 
temperatures but superconducting on further cooling) were found even just 
below $B_{\rm vg}^*$ in Bi/Sb films. According to our theory, this 
nonmonotonic behavior below $B_{\rm vg}^*$ is also understood as arising from 
the sum of the above-mentioned insulating $G_{fl}^{-1}$ above $B_m(0)$ and the 
superconducting $G_{\rm vg}^{-1}$ below $B_{\rm vg}^*$. This 
nonmonotonic behavior is visible because the window in $B_m(0) < B 
< B_{\rm vg}^*$ is moderately wide in contrast to that 
in MoSi case\cite{Okuma2}. Actually, the presence of a wide region $B_0 < B 
< B_{\rm vg}^*$, in which the resistance does not yield an activated behavior 
indicative of a vacancy- or interstitial-creep in a vortex solid, was pointed 
out in ref.7. As also mentioned in $\S 3$ of ref.19, this window should broaden with increasing disorder. However, our theoretical result shows that it is invalid to, based only on the nonactivated resistive behavior, identify\cite{Valles} the window in $B_m(0)$ (or $B_0$) $< B < B_{\rm vg}^*$ with a putative quantum liquid {\it phase} with low but nonzero resistance at $T=0$. 
Of course, in a system with still stronger disorder or at low enough $T$ (but 
above $T_{\rm rep}$), the contribution $G_n + G_{fl}$ is already negligible, and the leading term of $G_{\rm vg}(B=B_{\rm vg}^*)$, i.e., 
a (nonuniversal) constant $G \simeq G_{\rm vg}(B=B_{\rm vg}^*)$, should 
be observed as the total conductance, as actually seen in the state 1 
of ref.3 where a critical resistance value is much {\it larger} than $R_r$. 
On the other hand, the apparent critical resistance values in ref.7 seem to 
correlate to the corresponding $R_r$-values, implying that the samples in 
ref.7 have intermediate strengths of disorder. 

We emphasize again that, in the present theory, the intervening metallic 
behavior at a field is regarded not as a reflection of the true $T=0$ 
phase diagram but as a phenomenon in the intermediate (accessible) 
temperature range $T_{\rm rep} < T < T_{cr}^{\rm mf}$. This interpretation 
is never artifitial. In fact, 
a 2D FSI behavior was also observed in 2D-like but {\it bulk} (underdoped) 
YBCO,\cite{Chicago} in which the true $T=0$ critical behavior {\it must} be of 
a 3D VG type (see $\S 4$), and 
the system at nonzero temperatures behaves as if it have a 2D VG transition 
at $T=0$ (This situation is comparable with a dimensional crossover\cite{RI7} 
just above a {\it thermal} transition. Note that, in the present case, $\xi_{\rm vg}(B=B_{\rm vg}^*)$ diverges in $T \to 0$ limit). Similarly to this, in the present case the temperature variation of resistance curves in $B > B_{\rm vg}$ (see 
Fig.2) should become insulating at inaccessibly low temperatures {\it below} 
$T_{\rm rep}$, although the resistance curves 
in $B_{\rm vg} < B \ll B_{\rm vg}^*$ may decrease on cooling in $T 
> T_{\rm rep}$, more or less, as a result of classical (thermal) VG 
fluctuation. 

Now, the fermionic fluctuation term $\delta G_s$ of the 
conductance will be considered. Since discussing the normal part 
$G_n$ in details is beyond the scope of this paper, let us assume 
here $G_n(T>0)$ to show an almost metallic behavior. Then, 
the dynamics of the pair-field should be dominated by the dissipative term, and the {\it fermionic} fluctuation conductance $\delta G_s$, consisting of the 
Maki-Thompson terms and DOS terms, remains nonzero in low $T$ limit and was 
previously regarded as a correction to $G_n$ without examining in details 
(see $\S 7$ in ref.14). According to a recent systematic study in ref.25, 
$\delta G_s$ at $T=0$ is negative and given, on the Gaussian level, by 
$$R_Q \delta G_s (T=0) \simeq - {{\gamma^{(0)}(T=0)} \over {3 \pi \gamma}} {\rm ln}{1 \over {\mu(0)}}, \eqno(3.4)$$
which is valid above the mean field $H_{c2}(0)$ ($\mu(0) \simeq 
-1+B/H_{c2}^d(0) > 0$). Note that, 
in the dirty limit, the r.h.s. of eq.(3.4) is independent of material 
parameters. If the renormalization of LLL fluctuation is performed in terms 
of eq.(2.5) in deriving the corresponding one to eq.(3.4) applicable 
in $B < H_{c2}(0)$, one finds eq.(3.4) is replaced by 
$$R_Q \delta G_s^{(R)}(T=0) \simeq - {{2 \pi \gamma^{(0)}(T=0) r_B^2} 
\over {3 U_4}} |\mu(0)|,  \eqno(3.5)$$ 
where effects of pinning disorder on the fluctuation renormalization were 
neglected for simplicity. By using eqs.(2.28) and (2.29), 
the r.h.s. of eq.(3.5) becomes of the order of $- E_F \tau |\mu(0)|$. 
Further, the increase of $|\delta G_s|$ on cooling\cite{NMR} 
remains valid after the fluctuation renormalization is performed. 
Since $|\mu(0)| < 1$ within the GL theory, 
while $R_Q G_n \simeq E_F \tau$ in 2D and with no quantum correction,\cite{HF} 
the fermionic conductance $G_n + \delta G_s$ can significantly reduce with decreasing $B$ below $H_{c2}(0)$ due to the $|\mu(0)|$-dependence in eq.(3.5). This seems to explain the negative magnetoresistance (MR) 
in MoSi thin films, 
which was not visible in 3D-like films and in highly disordered 
films nonsuperconducting even at $B=0$.\cite{Okuma2} Further, a more remarkable negative MR had been also found in InO thin films\cite{HP,Gan2} showing the 
FSI behavior. In ref.34, the origin of this negative MR had 
been ascribed to a localization of bosons (i.e., pairs) which is also of a 
superconducting origin but, in contrast to our idea, does not seem to be 
supported through a microscopic calculation. We also 
note here that the data in ref.34 have suggested a {\it metallic} resistance in much higher fields than $B_{\rm vg}^*$ and even at low enough $T$. In our 
notation, this corresponds to a nonvanishing $G_n$ and is consistent with our 
assumption that, in contrast to ref.6, the FSI behavior should be explained 
based on a dissipative dynamics of the pair-field. 

Our scenario on the FSI behaviors is conventional and precludes a possibility 
of an intermediate metallic vortex phase at $T=0$. An argument favoring such an intermediate phase is based on the data suggestive of a quantum tunneling 
behavior in a ${\rm ln} G^{-1}$ v.s. $1/T$ plot\cite{Palo} and also on 
a computation of fluctuation coductivity applicable only to $B=0$ and 
based on a 
neglect of dissipative dynamics.\cite{DP} As verified in ref.4, however, 
such a metallic behavior over a broad field range 
was also seen in the case irrelevant to the vortex states, i.e., the case 
in a field parallel to the surface of thin film samples and with a current 
parallel to the field. This finding suggests that the the quantum tunneling 
behavior\cite{Palo} may not be due to an intrinsic origin. Actually, an intermediate metallic behavior tends to be seen in rather weakly disordered films with $R_r \leq 1$ k$\Omega$. As mentioned earlier, one needs to enter lower temperatures in a weaker disorder case in order to search a true low $T$ behavior. To the best of our knowledge, there is no {\it clear} evidence of an intermediate metallic behavior in a stronger disorder case. Actually, an intervening metallic 
phase was argued based only on models for the granular case in zero field 
which are inapplicable to the nonzero field case of homogeneous materials (see 
$\S 1$). Further, the argument\cite{DP} favoring $G(B=B_c) \simeq G_n$ 
certainly contradicts the $R_r$-dependence of critical resistance values 
measured in ref.3. 

\section {Discussions and Extension} 

In $\S 3$, we have discussed various resistive data suggestive of 
the FSI transition at $T=0$ by taking account of three different 
superconducting terms $G_{fl}$, $G_{\rm vg}$, and $\delta G_s$ of 
conductance. Those data have been explained by noticing that 
the experimentally accessible temperatures in real thin films with $R_r < R_Q$ 
will lie not in the vicinity of $T=0$ but within an intermediate temperature 
range defined by us. The resistance data in various samples 
are apparently incompatible with one another and are not explained 
comprehensively within a scenario taking account only\cite{MPAF} 
of $G_{\rm vg}$ without a detailed 
calculation. As in the thermal case,\cite{RI7,RI2} a couple of contributions 
to conductance are necessary in order to explain 
avalable data in a unified manner. Although we have explained, by 
focusing primarily on 
systems with relatively weak disorder, why the {\it apparent} critical 
resistance value tends to increase with increasing $R_r$, it has not been 
clarified whether the $G_{\rm vg}(B=B_{\rm vg}^*)$-value itself decreases or 
not with increasing $R_r$. 
Further, attention was not paid much to the values of 
exponents $z$ and $\nu$. Although they are usually assumed to take universal 
values, this assumption is not necessarily valid for a quantum transition in 
random systems. These issues are left for future studies. 

The extension of the present theory to 3D case will be explained here 
since, at a glance, recent data\cite{Okuma3} in thick (3D-like) MoSi films 
would seem to contradict the present theory. 
Similarly to the 2D case, the conductance $G_\perp$ for a current 
perpendicular to ${\bf B}$ can be seen as consisting of the four terms;
$$G_\perp = G_{n, \perp} + G_{fl, \perp} + \delta G_{s, \perp} + G_{{\rm vg}, \perp}, \eqno(4.1)$$
where the notation follows that in 2D case. Although the conductance defined 
in the applied field direction is not discussed here, our final conclusion remains valid for the parallel conductance. 
First, let us list the behaviors of each component in eq.(4.1) near $T=0$. 
In 3D, the normal 
part $G_{n, \perp}$ is safely assumed to be metallic, and the fluctuation 
correction $\delta G_{s, \perp}$ is nondivergent in contrast to 2D case 
and merely a small correction, insensitive to $B$ and $T$, to $G_n$, 
i.e., $\delta G_{s, \perp} \sim G_{n, \perp}$ O($1/E_F \tau$). 
According to ref.14, the AL fluctuation term $G_{fl, \perp}$ 
(except $G_{{\rm vg}, \perp}$), as in 2D case, approaches zero in $T \to 0$ 
limit above any transition field (see below). Further, it is easily found as a trivial extension of eq.(2.27) that the VG contribution $G_{{\rm vg}, \perp}$ has the form of a scaling function multiplied by $\xi_{\rm vg}^{-1}$, and 
thus, 
$$G_{\rm vg} (B_{\rm vg}) \sim \xi_{vg}^{-1}(T, B_{\rm vg}) \sim T^{1/z}, 
\eqno(4.2)$$
which implies $G_{{\rm vg}, \perp}(T \to 0, B_{\rm vg})=0$. Thus, above any 
transition field and at $T=0$, the total conductance is essentially 
equivalent to $G_n$. How about the $B$-$D$ phase diagram (Fig.3) at $T=0$, 
where $D$ measures the strength of {\it pinning disorder} ? 
In this case with dissipative 
dynamics, the quantum 3D GL model is equivalent to a classical 5D one, and the 
dimensionality of LLL fluctuation is three. Thus, the {\it ordinary} 
critical line $H_{c2}^*(T=0)$, signalling the onset of the ordinary long-ranged phase coherence, lies in nonzero fields in this 3D case, while the 
corresponding field in 2D is zero. 
Just as in the $B=0$ thermal transition in a 
bulk superconductor, the ordinary critical point 
$H_{c2}^*(0)$ (the left vertical line in Fig.3) 
is lowered from the mean field $H_{c2}(0)$ (the right vertical line in Fig.3) 
due to the 3D quantum fluctuation. 
When the pinning disorder is absent (i.e., $D=0$), as mentioned 
previously in $\S 5$ of ref.14, a first order 
vortex solidification transition at $T=0$ and in 3D should occur 
{\it above} $H_{c2}^*(T=0)$. Its position, the end point of 
$B_m(0)$-line, was indicated as an open circle in Fig.3. 
On the other hand, as found in ref.37, the {\it ordinary} superconducting 
transition in classical 5D case with nonzero $D$ is expected to be of second 
order. This fact may be useful in understanding the present situation 
due to the above-mentioned correspondence between the classical 5D 
and the quantum-dissipated 3D cases. Since the first order solidification at 
the open circle is also accompanied by the ordinary superconducting ordering, 
however, both transitions {\it must} connect with each other in $B$-$D$ 
diagram, as described by the chain line. 
Then, it is straightforwardly concluded that the 
resistance for any current direction {\it must} be zero at least at and below 
Max($B_m(0)$, $H_{c2}^*(0)$): 
According to the definition of VG ordering (see eq.(2.10)), the presence of the {\it ordinary} phase coherence inevitably implies the presence of VG ordering, 
although a VG ordering can generally occur\cite{RI2} with {\it no} 
ordinary phase 
coherence. Namely, if an estimated $B_{\rm vg}(D)$-line lies 
below $H_{c2}^*(0)$ for some $D$-values, then the resistance for {\it any} 
current direction vanishes not at $B_{\rm vg}$ but already at $H_{c2}^*(0)$. 
Thus, since there is no region with finite resistance 
below $H_{c2}^*(0)$, {\it no} metallic intermediate phase with much smaller 
resistance is possible in 3D 
and at $T=0$. Although, at large $D$, $B_{\rm vg}$ may lie 
above $H_{c2}^*(0)$, it is obvious that this conclusion is not affected. 
The resistance curves apparently residual on cooling, 
observed in ref.36, are the reflection not of a true $T=0$ metallic vortex 
phase but, just like the data in ref.33, merely of a pinning-induced 
enhancement of the vortex flow conductance in the vortex liquid {\it region} 
becoming narrower in $T \to 0$. 

\vspace{5mm}

\leftline{\bf Acknowledgement}

We thank S. Okuma for discussions.

\vfil\eject

\leftline{\bf Figure Caption}

\vspace{5mm}
Fig.1  

Examples of diagrams contributing to $G_{\rm vg}$. The solid curves denote the LLL fluctuation propagators, the chain curves are NLL propagators, the dashed lines are the pinning lines, and the hatched rectangle denotes $\chi_{\rm vg}$. 
\vspace{5mm}

Fig.2 

Details of 2D phase diagram near zero temperature. Here $H_{c2}(0)$ is the mean field upper critical field, $B_m(0)$ the $T=0$ vortex-solidification field 
estimated by neglecting the presence of the critical field $B_{\rm vg}$, the chain line denotes $T_{\rm rep}$, and the lower (upper) dashed straight line denotes $T_{cr}^{\rm mf}(B)$ ($T_{cr}(B)$). The quantum VG (QVG) critical regime is 
the $T$-dependent field range defined by eq.(2.15) around the {\it apparent} 
VG critical field $B_{\rm vg}^*$. Details of the true zero temperature limit 
below $T_{\rm rep}$ are not drawn. 

\vspace{5mm}

Fig.3 

Conjectured 3D $B$ v.s. $D$ ({\it pinning} disorder) phase diagram at $T=0$. 
The chain curve $B_m(0)$ and the open circle imply the first order transition accompanied by a melting of a vortex solid, the left vertical line 
is the ordinary superconducting transition line $H_{c2}^*(0)$, the right vertical line the mean field $H_{c2}$ line, and the curved portion in large $D$ of the solid line indicates $B_{\rm vg}$ occurring above $H_{c2}^*(0)$. The solid 
lines imply second order superconducting transition. 
As explained in the text, the VG susceptibility is 
divergent everywhere below $B_M \equiv$ Max($H_{c2}^*(0)$, $B_{\rm vg}$, 
$B_m(0)$) and hence, the resistance $R_\perp$ is zero in $B < B_M$, 
while $R_\perp \simeq G_n^{-1}$ in $B > B_M$. 

\begin{figure}[t]
\begin{center}
\leavevmode
\epsfysize=5cm
\epsfbox{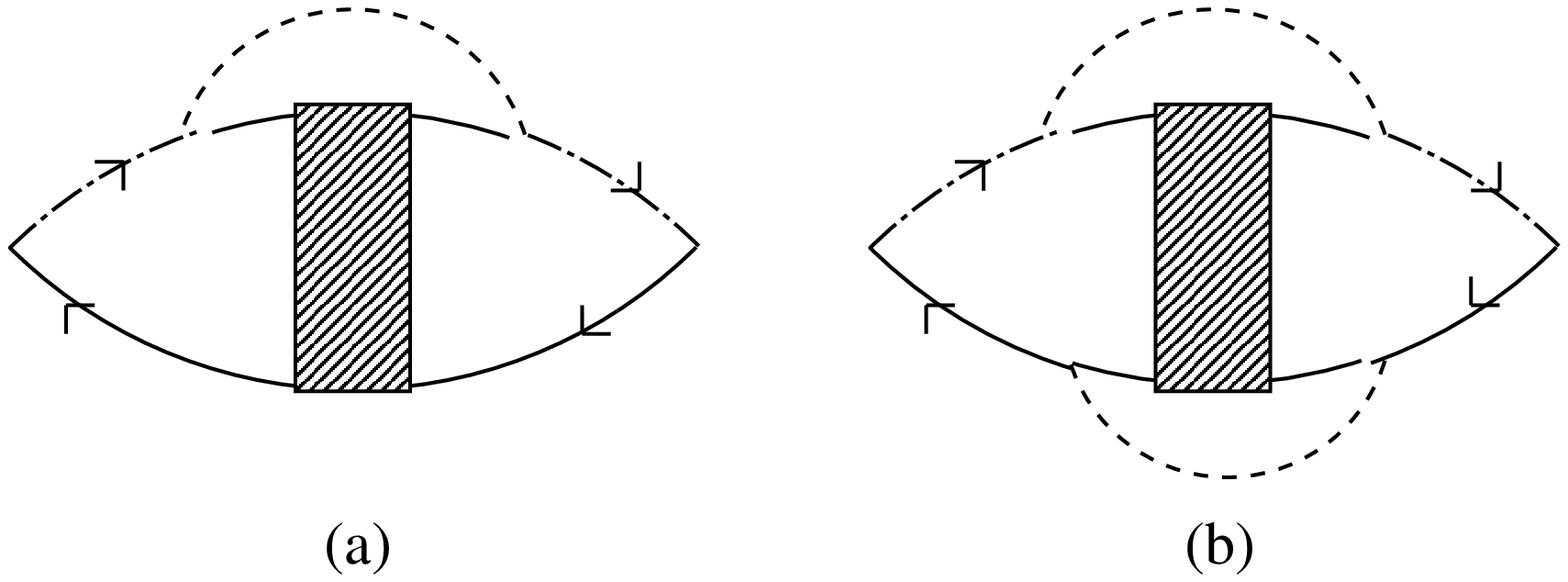}
\end{center}
\end{figure}

\begin{figure}[t]
\begin{center}
\leavevmode
\epsfysize=12cm
\epsfbox{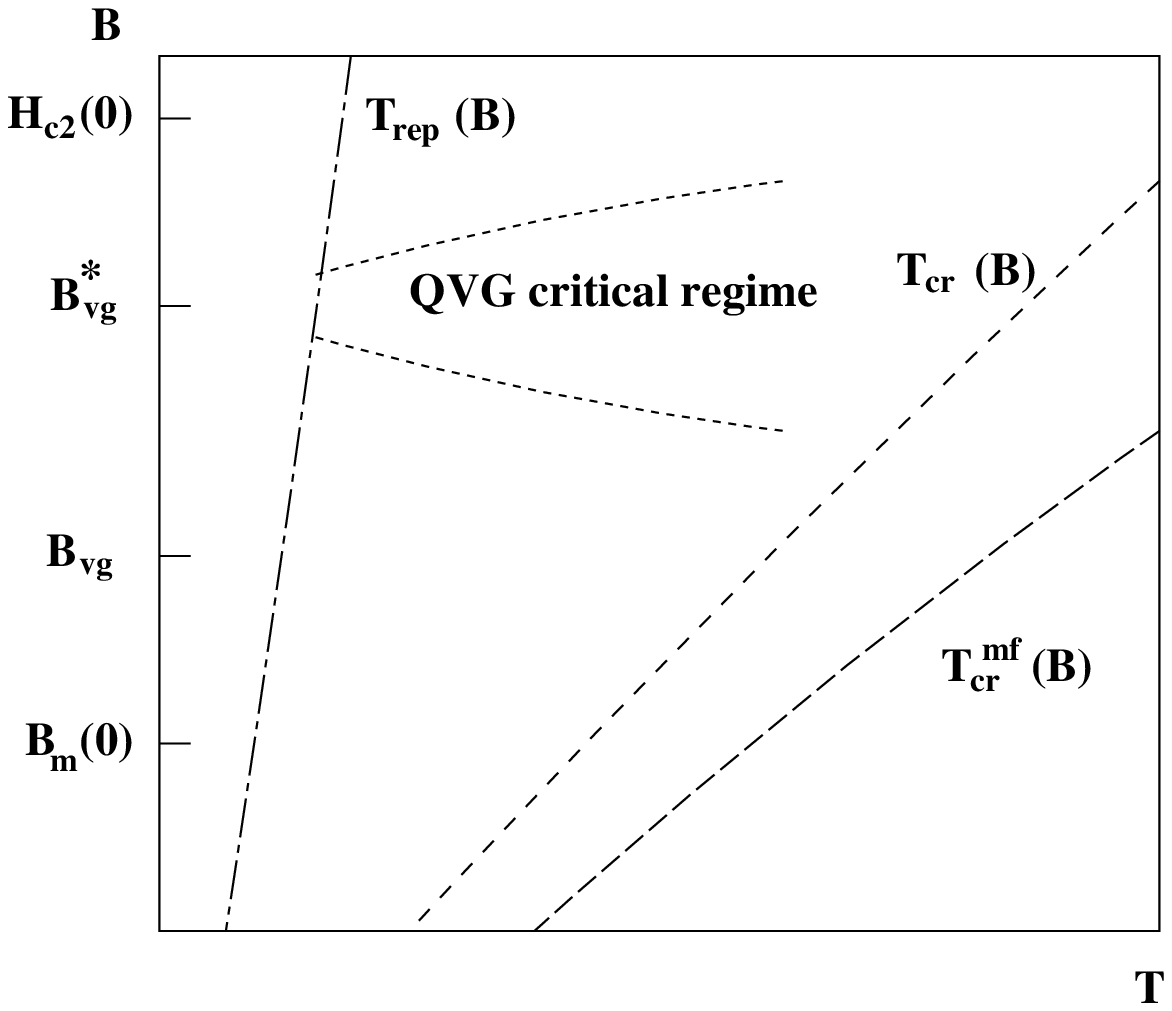}
\end{center}
\end{figure}

\begin{figure}
\begin{center}
\leavevmode
\epsfysize=10cm
\epsfbox{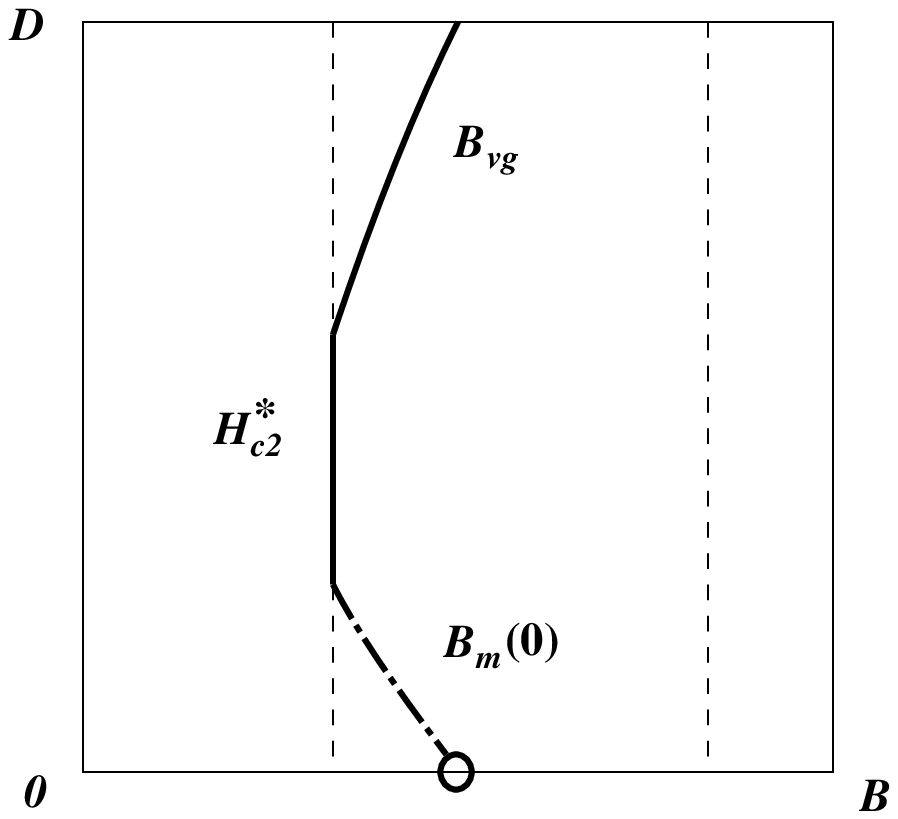}
\end{center}
\end{figure}


\begin{thebibliography}{99}
\bibitem{com2} Just like the pinning-free vortex solid, 
the static VG phase in high fields should be also described within LLL, and 
the NLL modes cannot participate in its description.
\end{thebibliography}
\end{document}